\documentclass[12pt]{article}

\title{Dynamics of Dollard asymptotic variables. Asymptotic fields
in Coulomb scattering}

\author{ G. Morchio and F. Strocchi \\ INFN, Sezione di Pisa, Italy}
\date{}

\makeglossary

\newtheorem{Theorem}{Theorem}[section]
\newtheorem{Definition}[Theorem]{Definition}
\newtheorem{Proposition}[Theorem]{Proposition}
\newtheorem{Lemma}[Theorem]{Lemma}

\usepackage{latexsym}
\usepackage[T1]{fontenc}
\usepackage{amsmath}
\usepackage{amssymb}

\def \endproof {\hfill \ensuremath{\Box}}

\def \ra {\rightarrow}

\def \AO {{\cal A}({\cal O})}
\def \AO' {{\cal A}({\cal O}')}

\def \Pf {{\bf Proof.\,\,}}

\def \be {\begin{equation}}
\def \ee {\end{equation}}
\def \ume {{\scriptstyle{\frac{1}{2}}}}
\def \ra {\rightarrow}

\def \eqq {\equiv}

\def \a {{\alpha}}

\def \d {{\delta}}
\def \eps {{\varepsilon}}

\def \l {{\lambda}}

\def \s {{\sigma}}

\def \ph {{\varphi}}

\def \A {{\cal A}}

\def \D {{\cal D}}

\def \H {\mbox{${\cal H}$}}
\def \J {{\cal J}}

\def \S {{\cal S}}

\def \U {{\cal U}}

\def \id {{\bf 1 }}

\def \d^nu {{\partial^\nu}}
\def \d^la {{\partial^\lambda}}
\def \d^o {{\partial^0}}

\def \abf {{\bf a}}
\def \bbf {{\bf b}}

\def \q {{\bf q}}

\def \p {{\bf p}}

\def \x {{\bf x}}
\def \y {{\bf y}}

\def \Rbf {{\bf R}}

\def\doppio#1{{\rm I}\kern-.1667em{\rm #1}}

\def\dR{{\rm \doppio R}}
\def\Q{\text{Q}\kern-.52em
    \text{\vrule height1.5ex width.5pt depth0pt}\kern.45em}

\def\dZ{{\mathchoice {\hbox{$\Ss\textstyle Z\kern-0.4em Z$}}
{\hbox{$\Ss\textstyle Z\kern-0.4em Z$}} {\hbox{$\Ss\scriptstyle
Z\kern-0.25em Z$}} {\hbox{$\Ss\scriptscriptstyle Z\kern-0.2em
Z$}}}}

\def\dC{{\mathchoice{\hbox{$\rm\textstyle\text{\kern.35em\vrule
   height1.5ex width.5pt depth0pt\kern-.35em C}$}}
{\hbox{$\rm\textstyle\text{\kern.35em\vrule
   height1.5ex width.5pt depth0pt\kern-.35em C}$}}
{\hbox{$\rm\scriptstyle\text{\kern.35em\vrule
   height1.5ex width.3pt depth0pt\kern-.35em C}$}}
{\hbox{$\rm\scriptscriptstyle\text{\kern.35em\vrule
   height1.5ex width.2pt depth0pt\kern-.35em C}$}}}}

\makeatletter

\@addtoreset{equation}{section}

\begin{document}

\maketitle

\begin{abstract}
Generalizing Dollard's strategy, we investigate the structure
of the scattering theory associated to any large time reference
dynamics $U_D(t)$ 
allowing for the existence of M{\o}ller operators.
We show that (for each scattering channel) 
$U_D(t)$ uniquely identifies, for
$t \to \pm \infty$,  
{\em asymptotic dynamics} $U_\pm(t)$; they are 
unitary {\em groups} acting on the scattering spaces,
satisfy the M{\o}ller  interpolation 
formulas and are interpolated by the $S$-matrix.
In view of the application to field theory models, we extend 
the result to the adiabatic procedure.
In the Heisenberg picture, 
asymptotic variables are obtained as LSZ-like
limits of Heisenberg variables; their time evolution is
induced by $U_\pm(t)$, which replace the usual
free asymptotic dynamics. 
On the asymptotic states, (for each channel) 
the Hamiltonian can by written in terms of the asymptotic variables as
$H = H_\pm (q_{out/in}, p_{out/in})$, $ H_\pm (q,p) $ the generator
of the asymptotic dynamics.
As an application, we obtain the asymptotic
fields $\psi_{out/in}$ in repulsive Coulomb scattering by an LSZ modified
formula; in this case, $U_\pm(t)= U_0(t)$, so that 
$\psi_{out/in}$ are \emph{free} canonical fields
and $H = H_0(\psi_{out/in})$. 
\end{abstract}

\vspace{20mm}

\vspace{2mm} 

\newpage
\section{Introduction}
As clarified by Dollard \cite{Do1}\,\cite{Do2}, 
long range interactions require a substantial modification
of the standard scattering theory and for potential scattering 
the situation is well understood \cite{De1}.

The possibility of exploiting Dollard's strategy, i.e., the use of a 
large time reference dynamics which is not free and not even a 
group, for controlling the infrared problem in 
QED is an interesting and, in our opinion, open
problem. 
Quite generally, one may ask what are the indications of Dollard strategy
for a Lehmann-Symanzik-Zimmermann (LSZ) condition for the 
construction of asymptotic charged fields.

An extension of Dollard approach to QED has been proposed by Kulish and 
Faddeev \cite{KF} and by Rohrlich \cite{JR}, involving both
the identification of a Dollard reference dynamics for large times, 
which takes into account the Coulomb distortion and the infinite 
photon emission, and the identification of the asymptotic fields 
with those proposed by Zwanziger and arising from generalized 
LSZ limits \cite{Zw}.


The outcome of that proposal represent a substantial departure from the 
standard framework; in fact, the time evolution of the asymptotic fields 
is not a group and the $S$-matrix is not invariant under it.
 
In Ref.\, \cite{JR}, such features arise as a consequence 
of the identification of the  
dynamics of the asymptotic fields with the Dollard modification of the
free dynamics; more generally, they stem from the incorporation 
of Coulomb and soft photons effects in the dynamics of asymptotic fields 
\cite{Zw}. 
It has also been argued that ``distorted'' asymptotic fields,
with a dynamics which is not a group, characterize the field theory version
of Coulomb scattering \cite{Sc2}.
This implies a substantial change from the LSZ and Haag-Ruelle (HR)
notion of asymptotic fields.


The aim of this note is to investigate 
\emph{the general structure of the scattering theory associated 
to a large time reference dynamics $U_D(t,s)$, satisfying the 
flow property (but not the group property), which 
allows for the existence of M{\o}ller operators
as strong limits}.

This will be done in a framework admitting different 
scattering channels, both in the Schroedinger and
Heisenberg pictures and also in the presence of
an adiabatic switching. 



The main result of Section 2 is that 
any reference dynamics leading to
the existence of M{\o}ller operators $\Omega_\pm$ as strong
limits for $t \to \pm \infty$ (with respect to a given
dynamics $U(t)$)
uniquely determines an {\em asymptotic 
dynamics} $U_\pm(t) $, which is {\em a strongly continuous 
unitary group and satisfies the standard interpolation formula,}
\be{  U(t) \Omega_\pm = \Omega_\pm U_\pm (t) \ . }\ee
Even if $U_D(t,s)$ is asymptotic to a free dynamics $U_0$ for large
$|t|$, namely $U_D(t+\tau,s) \sim U_0(\tau) \, U_D(t,s) $,
$ U_\pm (t) $ needs not to be free. 

The implications on the construction of 
Heisenberg asymptotic variables and $S$-matrix
are discussed in Section 3. The Heisenberg asymptotic 
variables, conventionally $q_{out/in}(t), p_{out/in}(t)$,
are covariant under $U(t)$ and, on the scattering space, 
the Hamiltonian can be written as
\be{ H = H_{out/in}(q_{out/in},p_{out/in}) \ ,}\ee 
where $H_{out/in}$ is the generator of  $U_\pm (t)$.
Thus, one recovers all the standard properties, with the only replacement
of the free Hamiltonian $H_0$ with $H_{out/in}$.
 
For of the extension of Dollard approach to the construction of
M{\o}ller operators in quantum field theory (QFT) models 
it is convenient to introduce an \emph{adiabatic regularization} 
and one must control the effects arising in its removal (Section 4). 
Quite generally, with a proper treatment of persistent effects
like mass renormalization counterterms, one obtains 
as above an asymptotic
dynamics which is a group and satisfies eq.\,(1.1) (Section 4).

In Section 5, the analysis of Sections 2, 3 is applied 
to the control of asymptotic fields
for (repulsive) Coulomb interactions; they are defined as strong 
asymptotic limits, are space-time covariant free fields
(with no Coulomb distortion) and are interpolated by the Heisenberg 
$S$-matrix, which is space-time translation invariant. 
In this case, the adiabatic procedure of Section 4
displays the factorization of infrared divergences. 
Thus, the Coulomb interaction 
and the corresponding distortion in Dollard reference dynamics do not imply
a distortion \cite{Sc2} of the asymptotic dynamics, 
which remains free.

An asymptotic dynamics which  is still a group, in accord with 
the results of Sections 2 and 4, but is not free, 
arises in QFT models with infinite photon emission. In fact, 
the present framework and analysis
apply to a model with realistic
photon emission, describing classical particles with Coulomb interaction and 
translation invariant coupling to the quantized electromagnetic field 
\cite{MS}.


\newpage
\section{Dollard asymptotic limits and asymptotic dynamics}
The important contribution by Dollard on Coulomb scattering 
has played a crucial role in the discussion
of scattering in the presence of long range interactions and 
of the infrared problem in quantum electrodynamics (QED) 
\cite{Do1}\,\cite{Do2}\,\cite{Si}\,\cite{De1}
\,\cite{KF}\,\cite{JR}.

\def \udt {U_D(t)}
\def \udtpm {U_{D \pm}(t)}
\def \udtmu {U_D(t)^{-1}}
\def \udts {U_D(t + s)}
\def \hdt {H_D(t)}
\def \ho  {H_0}
\def \opm {\Omega_{\pm}}
\def \omp {\Omega_+}
\def \omm {\Omega_-}
\def \li+ {\lim_{t \ra \infty}}
\def \lipm {\lim_{t \ra \pm \infty}}
\def \eht {e^{ i H t}}
\def \ut {e^{- i H t}}
\def \uas {U_{as}}

The main idea is that for the definition of the asymptotic limit one has to 
replace the free dynamics  by a ``distorted one'',  which is not a one-parameter 
group. 
In the interaction picture this amounts to replace $U_0(t) = e^{- i H_0 t}$  
by  $U_D(t) $, with \be{ \frac{dU_D(t)}{dt} = - i H_D(t)\, U_D(t),}\ee
where $\hdt$ is time dependent; for Coulomb scattering 
\be{ \hdt = H_0 + \frac{\a  m}{|\p| \, t},\,\,\,\,\,\,\,\,\udt = 
U_0(t) \exp{( -i \frac{\a m}{|\p|} \, {\rm sign} \, t\, \log |t|)}\, .}\ee 
The role of such a distortion is to allow for the existence of 
the strong limits
\be{ \Omega_{\pm} \eqq s-\lipm \eht \,\udtpm, }\ee
which do not exist for $e^{iHt} \, U_0(t)$, 
and of the S-matrix $S = \omp^*\, \omm$.

It is worthwhile to stress that even if $\udt$ is not a group, 
in the literature \cite{Sc2}\,\cite{JR}  it has been 
identified with ``the dynamics of the asymptotic states'', 
with the consequence  that time covariance  is inevitably lost. 
This is clearly  a serious obstruction
for the construction  of space time covariant asymptotic fields and in 
general for a LSZ approach. In fact, Zwanziger proposal \cite{Zw} 
for asymptotic fields in QED explicitly displays such a loss of covariance.
  
The aim of this Section is to re-discuss Dollard's strategy  and its 
implications in general, in order to question the above 
identification of the asymptotic dynamics and the loss of the group law.

To this purpose we shall start by analyzing the properties of the ``large time 
reference dynamics'' $U_{D \pm}(t,s)$ 
following merely by the fact that they  
allow for the existence of the M{\o}ller operators.

\def \udt  {U_D(t)}
\def \udts {U_D(t + s)}
\def \has  {\H^{as}} 
\def \hpm  {\H^\pm}
\def \limpm {\lim_{t \ra \pm \infty}}
\def \udtpm {U_{D \pm}(t)}
\def \udtspm {U_{D \pm}(t+s)}
\def \udtpma {U_{D \pm}^\a(t)}


\subsection{The group of the asymptotic dynamics}
We consider the scattering problem for the time evolution 
$U(t) = e^{ - i H t}$, in a Hilbert space $\H$, with respect 
to a ``large time reference dynamics'' defined by a family of 
unitary operators $\udt = U_D(t,0)$, such that,
for all ``scattering vectors'' $\psi$, 
\be{ || U(t)\psi - U_D(t) \ph|| \ra 0, \,\,\,\,\,\,
\mbox{for}\,\,\,\,t \ra \pm \infty.}\ee
 This means that for large $|t|$, $U(t)\psi$ is described by the 
reference dynamics $U_D(t)$ of the ``reference vectors'' 
$\ph$. In order to cover  the case of different scattering 
channels, indexed by $\a$ (e.g. corresponding to bound states), 
it is convenient to consider as reference vectors elements 
of {\em scattering spaces}  $\H^\pm_\a$,  
and reformulate the above convergence as the existence of 
\be{ \Omega_\pm^\a =  \limpm U^*(t) J_\pm^\a \,\udtpma \, ;  \ \ 
\Omega_\pm^\a: \hpm_\a \ra \H, }\ee
where $J_\pm^\a : \hpm_\a \ra \H$ are isometric operators describing
different channels, and $\udtpma$ are the reference 
dynamics in $\hpm_\a$ (see \cite{De1}).

Further  requirements for the choice of the large time reference 
dynamics $\udtpma$, in particular its relation with a free 
dynamics, will be discussed later. 
 
Such a formulation substantially reproduces the formulation of the 
scattering in the interaction picture, with the generalization 
in which the reference dynamics needs not  to be a group. 
In general, the states of $\hpm \eqq \oplus_\a \hpm_\a$ may be 
described in terms of ``scattering'' algebras $\A^\pm$ of 
operators acting in $\hpm$, respectively, and $\udtpm$ may be 
identified with functions of such variables 
(i.e., elements of the von Neumann closure of $\A^\pm$).

The operators $\opm^\a$ are automatically isometric; 
$\H_{(\pm)}^\a \eqq \opm^\a \H^\pm_\a$ are interpreted as the spaces 
of {\em asymptotic states} in $\H$  for $t \ra \pm \infty$ 
and will be assumed to be orthogonal and stable under time evolution
\be{ U(t) \,\opm^\a \H^\pm_\a = \opm^\a \H^\pm_\a.}
\ee
Property (2.6) follows from eq.\,(2.3) in the standard scattering theory, 
where the reference dynamics $\udt$ is a group. Its failure would lead to 
serious problems for the formulation of scattering theory (see below). 
\goodbreak

Until the end of Section 2.2, we shall work in a fixed channel 
and omit the index $\a$.

The fulfillment of eqs.\,(2.5), (2.6) is unaffected if $\udtpm$ is replaced 
by $\udtpm V_\pm$, with $V_\pm$  unitary operators; this amounts to 
replace $\Omega_\pm$ by $\Omega_\pm V_\pm$. 

\begin{Proposition}
Assuming eq.(2.5), 

\noindent
i) the following weak limit exist, $\forall s \in \Rbf$:
\be{ U_\pm(s) \eqq w-\limpm \udtpm^{-1}  
\udtspm =  \Omega^*_\pm U(s) \Omega_\pm \ ,}
\ee  
ii) the stability condition (2.6)
is a necessary  condition for any of the following  properties:
\newline \noindent 1) $U_\pm(s)$ are isometric operators, equivalently, the
limit in eq.(2.7) is strong;
\newline \noindent 2) $U_\pm(s)$, is a one-parameter group;
\newline \noindent 3) $U(s) \Omega_\pm = \Omega_\pm U_\pm(s).$
\end{Proposition}
\Pf\, In fact, $\forall \psi, \,\ph \in \hpm$, one has 
$$ (\psi, \,\udt^{-1}  \udts \ph) = 
(U^* (t) J \udt \psi, \,U(s) U^*(t + s) J \udts \ph) $$ 
\be{ \longrightarrow_{t \ra \pm \infty}  
(\Omega_\pm \psi, U(s) \Omega_\pm \ph) = (\psi, \, \Omega^*_\pm 
U(s) \Omega_\pm \ph).}\ee
Moreover, $U_\pm(s)$ isometric implies 
\be{ \id = U_\pm^*(s) U_\pm(s) = 
\Omega^*_\pm U(-s) \Omega_\pm \Omega^*_\pm U(s) \Omega_\pm}
\ee
and by multiplication by $\Omega^*_\pm$ and $\Omega_\pm$ on the right 
and on the left, respectively, one has, 
$\forall \psi \in \Omega_\pm \,\Omega^*_\pm \H \eqq P_\pm \H$,
\be{(U(s) \psi, \, P_\pm U(s) \psi) = (\psi,\, P_\pm \psi) = 
(\psi, \,\psi), }\ee 
i.e., $P_\pm U(s) \psi = U(s) \psi$; therefore $U(s)$ leaves $P_\pm \H$ 
stable. 

\noindent 
Similarly, if $U_\pm(s)$ is  a one-parameter group, one has 
$U_\pm(s)\, U_\pm(-s) =  \id $ and, 
since eq.\,(2.7) implies $U_\pm^*(s) = U_\pm(-s)$, $U_\pm(s)$ 
is unitary. Point 3) is immediate. \endproof

\vspace{1mm}
\noindent Quite generally, stability follows from any interpolation equation,
\be{ U(s) W = W V(s) \ \Rightarrow  \ U(s)\,W \H \subset W \H \, ;}
\ee  
however, given $U(t)$ and $W = \Omega_\pm $, 
$U_\pm(s)$ is the only possible solution of such an equation since 
$U(s) \Omega_\pm = \Omega_\pm V_\pm(s)$ implies $V_\pm(s) = U_\pm(s)$ 
by eq. (2.7).

\vspace{1mm}
 
Proposition 2.2 shows that stability under time evolution, eq.\,(2.6), implies 
that the limit (2.7) is strong; 
equivalently, that for  $t \ra \pm \infty$ 
\be{ U_{D\pm}(t+s) \sim U_{D\pm}(t)\,U_\pm(s),}\ee
with $U_\pm(s)$ a  one-parameter group,
{\em uniquely determined} by eq.\,(2.12). 
Here and in the following, $\sim$ denotes strong convergence 
to $0$ of the difference. 
\goodbreak

This means that for large $|t|$, the increments $t \ra t + s$ are described  
by the action of $U_\pm(s)$ on the right, briefly $U_{D\pm}(t)$ 
has the one-parameter groups $U_\pm$  as its ({\em unique})  
{\em right asymptotic group}.

For simplicity, we shall consider the case $t \ra \infty$, 
the extension to $t \ra - \infty$ being straightforward;
in most cases we shall omit the subscripts $\pm$, denote the corresponding 
{\em scattering spaces} $\H^\pm$ by $\H^\infty$ and $U_+$ by $U_{as}$.  

\def \d {\delta}  
\def \Has {H_{as}}
\begin{Proposition} Equations (2.5),(2.6) imply the existence of 
the following strong limit, 
defining a strongly continuous one-parameter group 
$\uas(s) = e^{ - i H_{as} s}$ ({\bf the asymptotic dynamics})
\be{ \udt^{-1}\,\udts \stackrel{s}{\longrightarrow}_{t \ra \infty}  \uas(s) .}
\ee
$\uas(s)$ is the unique 
solution  of eq.\,(2.12), for $t \ra \infty$, and satisfies 
the following {\bf interpolation formula}
\be{ U(s) \, \Omega = \Omega \, U_{as}(s) \, ,
\ \ \ \ \mbox{on}  \,\,\,\H^\infty.}\ee
\end{Proposition}

\begin{Lemma} If a sequence of isometric operators $U_n$ converges 
strongly,  for $n \ra \infty$, $U_n \stackrel{s}{\longrightarrow} \, \Omega$, 
then $\Omega$ is isometric and 
\be{ s-\lim_{n \ra \infty}  U_n^* \Omega = \id =\Omega^* \,\Omega \ ,}\ee
i.e.
\be{  s- \lim_{n \ra \infty} \,U_n^* = \Omega^*, \,\,\,\,\,\mbox{on} \,\,\,\Omega \,\H.}
\ee 
\end{Lemma}
\Pf\, In fact, $\forall \psi \in \H$, $U_n^* \Omega\,\psi = U_n^* (U_n \psi + \chi_n)$, 
with $\chi_n \stackrel{s} \ra_{n \ra \infty} 0$.
\endproof 

\vspace{2mm} \noindent 
{\bf Proof} (of Proposition 2.2). By eq.\,(2.5), $\forall \psi \in \H^\infty$
$$ \udtmu \udts \psi = \udtmu  J^* U(t + s)\, U(- t -s) J \udts \psi =$$ 
\be{ (U(t)^{-1} J \udt)^* U(s) (\Omega \psi + \chi(t)),}\ee
with $\chi(t) \stackrel{s} \ra 0,$ for $t \ra \infty$. \goodbreak 

\noindent By eq.\,(2.6) and Lemma 2.3, the right hand side converges to  
\be{\Omega^* U(s) \Omega \psi \eqq \uas(s) \psi,}\ee 
and, since by eq.\,(2.15) 
$\Omega \,\Omega^* = \id $ on  $ U(s)\,\Omega \,\H^\infty \subseteq \Omega \H^\infty$,  
$$\uas^*(s)\,\uas(s) = \Omega^* U(-s) \Omega\, \Omega^* U(s) \Omega = \id; $$ 
furthermore, $\uas(s)^* = \uas(-s)$ gives $\uas(s) \,\uas(s)^* = \id$.

\noindent Strong continuity follows from the definition of $\uas(s)$, eq.\,(2.18); 
moreover, since
$\Omega \,\Omega^* = \id $ on $ \Omega\, \H^\infty$, 
eq.\,(2.6) implies the group law and eq.(2.14), 
the latter by multiplying eq.\,(2.18) by $\Omega$. \endproof  

\def \uast {U_{as}(t)}

\vspace{2mm}

\noindent {\bf Remark 1.} 
Given eq.\,(2.5), by Prop.\,(2.1) strong convergence of the l.h.s. 
of eq.\,(2.13) is actually equivalent to eq.\,(2.6).
Equation (2.14) shows that the time evolution of the 
asymptotic states in $\H$, 
i.e. of the vectors in $\Omega\, \H^\infty$,  is given by a 
\emph{one-parameter group} $\uast$, uniquely determined by $\udt$. 
This property, which holds for Coulomb scattering 
with $H_{as} = H_0$ \cite{Do2}, is therefore much more general, 
only depending on the existence of the asymptotic limits, eq.(2.5), 
and on the stability of the scattering states under the time evolution, 
eq.(2.6). 

\vspace{1mm}
\noindent {\bf Remark 2.} A free dynamics $U_0(t)$ has not entered in the above 
analysis and in particular $\uast$ needs not to be a free dynamics. 
On the other side, the relevance of $\uast$ is clearly displayed by 
eq.(2.14), which gives it uniquely in terms of the M{\o}ller operators. 
Moreover,  if $V(t)$ is a family of unitary operators, 
any interpolation formula 
\be { U(t) W = W V(t), \ \ \ \mbox{with} \ \  
  W^* \,W = \id \, , \ \ \ \mbox{on} \ \ \H^\infty }
\ee
implies that $W \H^\infty$ is stable under $U(t)$ and
$V(t)$ is a one-parameter group. Therefore,  
$\udt$ cannot be a candidate for such an interpolation with respect to $U(t)$, 
 (contrary to statements appeared in the 
literature \cite{Sc2}), unless it has the group property.

\vspace{1mm}
\noindent {\bf Remark 3.} The above results are relevant for the construction 
of asymptotic limits in models with soft  photon emission.
In fact, as a consequence 
of the absence of charged one-particle states with a definite energy dispersion 
law $E(\p)$ (as implied  by the presence of an infinite number of 
asymptotic photons), the spectrum of $H$ on the scattering states cannot 
be that  of the free Hamiltonian $H_0$ and therefore eq.\,(2.14) cannot 
hold with $H_{as} = H_0$. 
This is explicitly shown by the model of Ref.\,\cite{MS}.

\noindent {\bf Remark 4.} 
It is worthwhile to remark that 
$\tilde{U}_D(t) \eqq \udt V$ 
gives 
\be{ \tilde{U}_{as}(t) = V^* \uast V, \,\,\,\,\,\tilde{\Omega} = \Omega\,V.}\ee
This occurs if a change of initial time $U(t) \ra U(t -s)$ 
is accompanied by $U_D(t) \ra U_D(t) U_D(s)^* = U_D(t, s)$ 
(see Appendix A) rather than  by $U_D(t) \ra U_D(t-s).$
In this case, the asymptotic dynamics, the corresponding M{\o}ller operators 
and the $S$-matrix are given by 
$$U_D(s)\, \uas(t)\, U_D(s)^*, \,\,\,\,\,\, \Omega(s) = U(s) 
\Omega(0) U_D(s)^*,$$
\be{ S(s) = \Omega_+^*(s) \Omega_-(s) = U_D(s) S(0)\,U_D(s)^*.}
\ee 
\goodbreak

\def \uint  {U_-(t)}
\def \uoutt {U_+(t)}
\def \uout  {U_+}
\def \udtst {U_D(t + s,t)}
\def \dhts  {\delta H_D(t +s)}


\subsection{Asymptotically free reference dynamics }

Up to now, $\udt$ 
has been  only constrained  to satisfy 
eqs.\,(2.5), (2.6). In many cases,, however, $\udt$ is constructed as a 
modification of the free dynamics $U_0(t) \eqq e^{ -i H_0 t}$, 
needed when $U(t)^{-1} J U_0(t)$ does not converge for $t \ra \infty$. 
If the corrections to the free Hamiltonian are small for
large times, in a suitable sense, interesting relations arise
between $\udt$ and $U_0(t)$.
In Appendix A, a 
notion is given of  ``asymptotically vanishing'' 
modification of the free Hamiltonian, 
which implies (Proposition A.3) the following
relation between $\udt$ and $U_0(t)$.

\begin{Definition} 
$U_D(t)$ is said to be {\bf asymptotically free} if it has 
$U_0$ as a left asymptotic group,
\be  U_D(t+s) \sim_{|t| \ra \infty} U_0(s) \, \udt ,
\ee
uniformly for $s$ in finite intervals.
\end{Definition}

\vspace{1mm}
\noindent
Eq.\,(2.22) means that, for large $|t|$,
the increments $t \ra t +s $ are described by the action of $U_0(s)$ 
{\em on the left}. 
Contrary to the case of a right asymptotic group (eq.\,(2.12), 
eq.\,(2.22) does not identify a unique left asymptotic group; 
in particular, it is always satisfied by $U(t)$, as 
a consequence of  eq.\,(2.5). 
Therefore, eq.\,(2.22) should not be viewed as a reconstruction of
$U_0$ from $\udt$, but rather as \emph{a constraint on} $\udt$, 
 \emph{given} $U_0$.

The most important consequence of the above notion is 
the following.

\begin{Proposition} If  $U_D(t)$ is  asymptotically free, then 
\be{\uas(s) = s-\lim_{t \ra \infty} \udt^{-1} U_0(s) \udt,}
\ee 
and  $\s(H_{as}) \subseteq \s(H_0)$. 
\end{Proposition}
\Pf\, Eq.\,(2.23) immediately follows from eq.\,(2.13) and eq.\,(2.22). 
Uniform convergence for $s$ in finite intervals implies convergence 
in the sense of tempered distributions.

\noindent Moreover, $\forall f \in \S(\Rbf)$, denoting by $\tilde{f}$ 
its Fourier transform,  
$$ \uas(f) = \int_{\s(H_{as})} dE_{as}(\l) e^{- i \l s} f(s) d s = $$ 
$$= \int_{\s(H_{as})} 
dE_{as}(\l) \tilde{f}(\l)= \lim_{t \ra \infty} \udt^{-1}\int_{\s(H_0)} dE_{0}(\l) 
\tilde{f}(\l) \,\udt,$$ and, therefore, the integral vanishes if 
supp\, $\tilde{f} \cap \,\s(H_0) = \emptyset$, 
which implies $\s(H_{as}) \subseteq \s(H_0)$. \endproof

\vspace{2mm}
\noindent 
{\bf Remark} Even if $U(t)^{-1} J U_0(t)$ does not converge,  
$U(t)$ may satisfy
$$U_0(t+s)^{-1} J^* U(t+s) \sim U_0(t)^{-1} J^* U(t),\,\,\,\,
\mbox{on}\,\,\, \Omega \,\H^\infty,$$
equivalently
\be{ J^*\,U(t+s) \Omega \sim  U_0(s) J^* U(t)\,\Omega,}\ee
i.e., $J^*\,U(t)\,J$ may have $U_0 $ as a left asymptotic group. 

\noindent Given eq.\,(2.5),
eq.\,(2.24) is actually equivalent to eq.\,(2.22); 
in fact, by eq.\,(2.5), on $\H^\infty$ ($J^*\,J = \id$)
$$ J^* U(t+s) \Omega \sim U_D(t + s),$$ 
$$U_0(s) U_D(t)= U_0(s)J^* U(t) U(t)^{-1} J U_D(t) \sim U_0(s) J^* U(t) \Omega.$$
Then, both the left and the right hand sides of eqs.\,(2.22), 
(2.24) are asymptotically equivalent and  $\udt$ 
is asymptotically free iff so is $U(t)$
on $ \Omega \,\H^\infty$. 


\subsection{Time reversal invariance and S-matrix}
The above analysis of  the scattering for each channel $\a$ allows for the 
description of the scattering in terms of vectors in $\H^\pm = \oplus_\a 
\H^\pm_\a$ 
and of operators $\Omega^\pm = \oplus_\a \Omega^\a_\pm$, $U_\pm(t) = 
\oplus_\a U_\pm^\a$: $\forall \psi \in \Omega_\pm \H^\pm$ 
\be{ U(t) \psi \sim_{t \ra \pm \infty} \sum J_\pm^\a \udtpma \psi_\pm^\a, \,\,\,
\,\psi_\pm^\a = \Omega^{\alpha\, *}_\pm \psi \in \hpm,}\ee 
\be{U(t) \Omega_\pm \psi_\pm = \Omega_\pm\,U_\pm(t)\,\psi_\pm,}\ee
with  $U_\pm(t)$ the one-parameter groups which describe the time evolution 
in $\hpm$.

Under the above assumptions, eqs.\,(2.5), (2.6), the $S$-matrix 
$S = \Omega_+^* \Omega_-: \H^- \ra \H^+$ satisfies 
\be{ S \,U_-(t) = U_+(t) \,S,}\ee
and is isometric iff \be {\Omega_- \H^- = \Omega_+ \H^+.}\ee   

In general, under a modification of the reference dynamics by 
unitary operators $V_\pm$
the $S$ matrix transforms covariantly and eq.\,(2.28) still 
holds with $U_\pm$ redefined as in eq.\,(2.20):
\be{\udtpm \ra \udtpm \,V_\pm\, , \ \ \ S \ra V_+^*\,S\,V_-.}\ee

Up to now, the description of the scattering 
spaces and the associated operators $J_\pm$, $U_{D\pm}$ are independent and 
unrelated and one has two asymptotic dynamics $U_{as} (t)$, $as = \pm$, 
acting in $\hpm$.  
A link between them may be obtained by using time reversal invariance. 
In particular, one may look for conditions which allow for an identification 
of the two asymptotic dynamics at $t \ra \pm \infty$. 
To this purpose, we assume that\goodbreak

\noindent 
1) the Hamiltonian $H$ is invariant under time reversal $T$;

\noindent 
2) the operators $J_\pm^\a: \H_\alpha^{\pm} \ra \H$ may be chosen  such that 
\be{ \mbox{Ran} J_-^\a =  \mbox{Ran} J_+^\a= \, T\,\mbox{Ran} J_-^\a.}\ee

\noindent 
3) The invariance of $H_+$ under the time reversal 
$T_+ \eqq \oplus_\a J_+^{\a\, *}\,T\, J_+^\a$ 
induced by $T$ on $\H^+$, i.e.
\be{ H_+ = T_+\,H_+\,T_+}
\ee

\vspace{1mm}
Given  $U_{D+}^\a(t)$ for $t > 0$, satisfying eqs.\,(2.5), (2.6), 
1) and 2) imply that one may always take, for each channel  
(omitting  the index $\a$, for brevity)
\be{U_{D -}(-t) = J_-^* T J_+ U_{D +}(t) J_+^* T J_- = 
T_\infty^{-1}\,U_{D +}(t) \, T_\infty,\,\,\,\,\,T_\infty \eqq J_+^* T J_-,}
\ee
as a reference large time dynamics, for negative times. 
Then, the so defined reference dynamics 
satisfy eqs.\,(2.5), (2.6) for both $t > 0$ and $t < 0$  
and
\be{\Omega_- \eqq \lim_{t \ra \infty} U(-t)^{-1}\,J_-\, U_D(-t) = T\, 
\Omega_+ T_\infty \, .}
\ee
Furthermore, with $T_\infty \eqq \oplus_\a J_+^{\a\, *}\,T\, J_+^\a  $, 
one has
\be{\uint = T_\infty^{-1} \uout(-t) T_\infty, \ \
\mbox{i.e.,} \ \  H_- = T_\infty^{-1} H_+ T_\infty \, .}
\ee  

\def \smatrix {$S$-matrix\, }
\def \sh  {S_H}

\noindent
3) is equivalent to 
\be{  H_- = \J^* \,H_+ \,\J, \,\,\,\,\J \eqq \oplus_\a J_+^{ \a \,*}\, J_-^\a : 
\H_- \ra \H_+, }
\ee
corresponding, for each channel, to the equation 
$$ H_- = T_\infty^{-1} H_+ T_\infty = J_-^* T J_+ H_+ J_+^* T J_- = 
J_-^* J_+ T_+ H_+ T_+ J_+ J_-^* = $$ 
$$= J_-^* J_+\, H_+\, J_+^* J_- \, .$$ \goodbreak

\vspace{1mm}
In conclusion, if eq.\,(2.30) holds, one may identify
\be{\H^+ = \H^- \eqq \H^{\infty}, \,\,\,\A^+ = \A^-\eqq \A_\infty, \,\,\, H_+ = 
H_-\eqq H_\infty.}
\ee
Under the above assumptions, one has
\be{ T_\infty \,S \,T_\infty = S^*, \,\,\,\,\,\mbox{on}\,\,\,\,\H^\infty,}\ee
and 
\be{S\, U_\infty(t)\,=\, U_\infty(t) \,S \ , \ \ \ \ \  \ 
U_\infty(t) \eqq e^{-i H_\infty t } \ .}
\ee 
The resulting picture is strictly analogous to standard scattering theory, 
with the one-parameter group $U_\infty(t)$ playing exactly the same role of 
the free dynamics $U_0(t)$, on the scattering spaces.


\section{Scattering in the Heisenberg picture}
In view of the possible extension of the above discussion to quantum 
field theory, where the Schroedinger picture  meets substantial 
problems, we discuss the implication of the above results 
on the formulation of scattering in the Heisenberg picture.

\subsection{Heisenberg asymptotic variables}

\def \ahas {A^H_{as}}
\def \Aio {A^H_{out/in}}
\def \has {\H_{as}}
 
The discussion of the asymptotic limits 
$t \ra \pm \infty$ and of the scattering 
processes may  more economically be done in terms of 
(Heisenberg) asymptotic  variables   acting in the 
Hilbert space $\H$.

For each variable $A \in \A^\pm$ 
we introduce a {\em Heisenberg asymptotic variable} $A^H_{out/in}$
{\em acting on} $\H_{out/in} \eqq \Omega_\pm \H_\pm \subseteq \H$, 
by the equation 
\be{ (\psi, \,A^H_{out/in}\,\psi) = 
(\psi_\pm, \, A \psi_\pm)\, , \ \ \psi \in \H_{out/in} \, ,
\ \ \ \psi_\pm = 
\Omega^*_\pm \psi \in \hpm \, , }\ee
i.e.,
\be{ \Aio = \Omega_\pm \,A\, \Omega^*_\pm,\,= \oplus_\a \Omega_\pm^\a A^\a 
\Omega_\pm^{\a \,*} \eqq \oplus_\a  
A_{out/in}^{H\,\a} \ \ \ \ \ \mbox{in} \ \ \ \H_{out/in} \, .}\ee
The operators $\Aio$ have the same (canonical) structure of 
the original $A \in \A^\pm$ and, by eq.\,(2.14),
their Heisenberg time evolution is 
\be{ U(t)^*\,A^H_{out/in} U(t) = 
\Omega_\pm \,U_\pm(t)^*\, A\, U_\pm(t) \Omega_\pm^*
 = (U_\pm(t)^*\, A\, U_\pm(t))_{out/in} \  .}\ee 
Eq.\,(3.3) gives the time evolution of the Heisenberg asymptotic
operators in terms of the asymptotic dynamics defined by eq.\,(2.13)
in the scattering spaces $\H_\pm$.
The  operators $\udtpm$ are assumed to be functions of 
operators belonging to $\A^\pm$, i.e. to belong to their 
von Neumann closure. 
Therefore, by eq.\,(2.13),  
so are the operators $U_\pm(t)$ and their generators $H_\pm$ may be 
identified with functions $H_{out/in}$
of such variables; denoting 
for convenience by $q_\pm, p_\pm$ the generators of $\A_\pm$, 
\be{H_\pm  = H_{out/in}(q_\pm, p_\pm).}\ee
Thus, by eq.\,(2.14), on $D(H) \cap \H_{out/in}$ one has
 \be{ H =\Omega_\pm H_{out/in}(q_\pm, p_\pm) 
\Omega^*_\pm =  H_{out/in}(q^H_{out/in}, p^H_{out/in}). }
\ee
i.e., $H$ is given, on $\H_{out/in}$ , by a function
of the Heisenberg asymptotic operators, the function being 
identified by the
Dollard dynamics through the construction of Proposition 2.2.
 
A relevant question is whether $\Aio$ may 
be obtained exclusively in terms of Heisenberg operators 
$A^H$ acting in $\H$.
Indeed, by putting, for each channel $\alpha$, $ A \in \A_\pm$,
\be { \alpha^t_{D\pm}(A)\,\eqq U_{D \pm}(t) \,A \, U^*_{D \pm}(t)\ , 
\ \ \ \ \ \ A^H \eqq J_\pm \,A\,J_\pm^{-1} \, ,}
\ee
one has, on $ \H_{out/in}$,
\be{\Aio = s-\limpm U(t)^* \,(\alpha^t_{D\pm}(A))^H \, 
U(t)  }
\ee
\be{ = s-\limpm U(t)^* \, U^H_{D \pm}(t) 
\,A^H \, U^H_{D \pm}(t)^* \,  U(t)  }
\ee
with $ U^H_{D \pm}(t) \eqq J_\pm \, U_{D \pm}(t) \, J_\pm^{-1} $. 

Eq.\,(3.8) gives the Heisenberg asymptotic operators as 
strong limits of Heisenberg Dollard-evolved operators. 
We recall that, in the standard Heisenberg
scattering theory, the role of the free evolution is twofold:
i) it provides the reference large time dynamics for
the existence of the asymptotic limits, ii) it describes the 
time evolution of the resulting (Heisenberg) asymptotic fields.
It should be stressed that a time dependent modification of 
the free dynamics (needed for the existence
of the asymptotic limit) cannot describe the time evolution
of the asymptotic variables, since it is not a group; on the other hand
it uniquely determines \emph{the group of asymptotic dynamics},
eq.\,(2.13), which gives the 
time evolution of the Heisenberg asymptotic operators, eq.\,(3.5).




\subsection{The S-matrix}
The  definition of the \smatrix in Section 2.3 essentially relies on the 
interaction picture and for a comparison with the 
discussion of scattering in field theory 
in the Heisenberg picture, it is convenient to introduce  the \smatrix $S_H$ 
in the Heisenberg picture. 

Under the standard assumption (2.30),  
$S_H$ is defined on $\H_{in} =\Omega_- \H_- \subseteq \H$ by 
\be{ S_H \eqq \Omega_+ \,\J\,\Omega_-^* = \Omega_-\,S^*\,\J\,\Omega_-^*,}\ee 
and it is unitary iff Ran\,$\Omega_- = $ Ran \,$\Omega_+$.

An important property of $S_H$ is its invariance under the time evolution 
$U(t)$, which follows from eq.\,(2.35),(whereas, under the above 
assumptions the $S$-matrix $S$ in the Schroedinger  picture is invariant under 
$U_{\infty} $, eq.\,(2.38)). 
In fact, by using eqs.\,(2.14), (2.35), one has 
$$ U(t)\,S_H = U(t) \Omega_+ \,\J\Omega_-^* = 
\Omega_+ \,U_+(t)\, \,\J\,\Omega_-^*= $$
\be{= \Omega _+\,\J\,U_-(t) \,\Omega_-^* = \Omega_+ \,\J\, \Omega_-^* U(t) = 
S_H\,U(t).}
\ee
Therefore, with the identifications (2.36), one has 
\be{ S_H = \Omega_+\,\Omega_-^* = \Omega_- 
\,S^* \,\Omega_-^*, \,\,\,\,U(t)\,S_H = S_H\,U(t).}
\ee
Furthermore, $\sh$ interpolates between 
the Heisenberg $out/in$ asymptotic variables; in fact, 
by  eq.\,(3.2), $\forall \a$, one has:
\be{  A_{out}^{H\,\a} = \sh\, A_{in}^{H\,\a}\,
\sh^*, \,\,\,\,\,\mbox{on}\,\,\,\,\H_{out} \subseteq \H.}\ee
As a result,  in each irreducible representation 
of the operators $A^H_{out/in}$, as also emphasized in the
LSZ and Haag-Ruelle approach, \,$\sh$ is identified by eq.\,(3.12), 
so that it may be obtained exclusively in terms of the Heisenberg 
asymptotic variables $\Aio$,  given by limits of 
Heisenberg variables, eq.\,(3.7), with no reference to the M{\o}ller 
operators, nor to an interaction picture.

\vspace{2mm} Eq.\,(3.7) has the form of an asymptotic LSZ (HR) formula,
where the free evolution (usually encoded in the test functions)
is replaced by the Dollard transformation $\alpha_{D \pm}^t$, eq.\,(3.6):
\be{\Aio = s-\limpm A^H_{LSZ}(t)  \ \ \ \ \ \ 
\mbox{on} \ \,  \H_{out/in} \, ,  }
\ee
$A^H_{LSZ}(t) \equiv U(t)^* \,(\alpha^t_{D\pm}(A))^H \, U(t)$.
Thus, the generalized Dollard strategy provides a candidate for 
a modified LSZ construction of asymptotic variables and 
\smatrix, with the following results:

\noindent 1) asymptotic variables (or fields) are obtained as strong 
limits of Heisenberg variables, eq.\,(3.13) playing the role of the LSZ 
asymptotic condition; they are time translation covariant and 
their time evolution is given by the asymptotic dynamics $U_\pm $, eq.\,(3.3);

\noindent 2) the Hamiltonian $H$ may be written as a function 
$H_{out/in}$ of the asymptotic variables, eq.\,(3.5); in contrast to 
ordinary scattering theory, in general $H_{out/in}$ is not a free
Hamiltonian. However, if $ U_{D}(t) $ is asymptotically free, 
eq.\,(2.22), $H_{out/in}$ is related to the free Hamiltonian 
by eq.\,(2.23). In fact, in a model of QED \cite{MS},
$H_{out/in}$ is a free Hamiltonian in terms of the ordinary
LSZ limit of the photon field and of the modified LSZ
limit of the charged fields.
 
\noindent 3) the \smatrix\,intertwines between the $in/out $ variables, 
eq.\,(3.12), and this property defines it up to a phase in each irreducible 
representation of the asymptotic variables.

\vspace{1mm} 
The emerging picture is very different from that proposed by 
Schweber, Rohrlich and Zwanziger \cite{Sc2}\,\cite{JR}\,\cite{Zw}, 
since there the evolution of their  
asymptotic states and asymptotic fields is given
by the Dollard-Kulish-Faddeev reference dynamics $U_D(t)$, which is 
not a group (with the drawback that the \smatrix in the 
Schroedinger picture is not covariant under it and even 
energy conservation becomes problematic).
In their approach, the absence of an interpolation formula
also prevents the expression of the Hamiltonian $H$ as a function 
$H_{out/in}$ of the asymptotic fields, eq.\,(3.5).

The point is that the asymptotic fields, $ A^Z_{out/in}(t)$, 
proposed by Zwan\-ziger actually describe a large time behavior 
of the fields and are {\em not} the result of (strong) LSZ 
asymptotic limits, in contrast to eq.\,(3.13).


\section{Adiabatic procedure for Dollard reference dynamics}
The standard regularization of the dynamics at large times by an 
adiabatic switching  
is not a substitute of Dollard strategy \cite{Do3}. 
However, its use 
{\em in combination} with  a reference dynamics $U_D(t)$ provides 
useful information, already in the case of Coulomb scattering, 
as we shall see. 

Moreover, in general, its use is necessary 
for the construction of M\"{o}ller operators in quantum field theory models, 
in particular for infrared models.
In fact, it use in the model of Ref.\,\cite{MS} allows
for the full control of the infrared problem, including the
asymptotic limit of the charged fields.

\def \ueps  {U^\eps}
\def \uepst {U^\eps(t)}
\def \uepsd {U_D^\eps}
\def \uepsdt {U_D^\eps(t)}
\def \Omepm  {\Omega^\eps_\pm}
\def \epst  {e^{- \eps |t|}}
\def \Omze {\Omega_0^\eps}
\def \Omz {\Omega_0}
\def \limp {\lim_{t \ra \infty}}
\def \veps {\tilde U^\eps}
\def \vs {\U(s)}

\vspace{1mm}
The standard adiabatic procedure consists in switching off the 
interaction for large times by
replacing the coupling constant, $g \mapsto e^{-\eps |t|} g$, i.e., 
$H_I(g) \mapsto H_I(e^{-\eps |t|} g)$. 
This amounts to the replacement of $U(t)$ by $U^\eps(t)$ with 
$U^\eps(t)$ satisfying
\be{i d U^\eps(t)/d t = 
(H_0 + H_I( e^{ - \eps |t|} g)) \,\uepst 
\eqq H^\eps(t)\,\uepst\, , \ \ \ U^\eps(0) = \id \, .}
\ee
In the treatment of scattering processes of $N$ particles, such a 
procedure applies to
the $N$ particle channel, for which $\H_\pm$ can be identified with $\H$.\goodbreak

Quite generally, an adiabatic switching can be formalized
 by the replacement of 
$U(t) = U(g,t)$ by $U^\eps(t) = U^\eps(g,t)$ satisfying

\noindent
i) $U^\eps(t) \to  U(t)$,  for $\eps \to 0$,

\noindent
ii) $U^\eps(g, t+s) = U^\eps(g e^{-\eps s},t) \,  U^\eps(g,s)$ , for sign$\,s =$ 
sign$\,t$.

\vspace{1mm} \noindent
Eq\,(ii) follows if $U^\eps(g,t)$ 
is the unique solution of eq.\.(4.1) and 
naturally arises from the time ordered exponential formula for its solution.
 
In order to identify the non-trivial points arising for the removal of the 
adiabatic switching
after the limit $ t \ra \pm \infty$, we first discuss the case of 
scattering with respect to a large time reference 
dynamics $U_0(t)$ which is a one-parameter unitary group in $\H$, with 
the following assumptions:

\noindent {\bf 1}) {\em 
Existence of the M{\o}ller operators for $\eps > 0$}. Namely,
$U^\eps(g,t)^{-1} \, U_0(t)$  converge strongly, to isometric operators 
$\Omze(g) \eqq \Omze$, 
\be{s-\limp \,\uepst^{-1} U_0(t) = \Omze(g) \, .}
\ee     
By the same argument of Proposition 2.2, with $U, U_D$ replaced 
by $U_0, U^\eps$ respectively, eq.\,(4.2) implies that the following limit 
exists and defines a unitary one-parameter group 
$\veps(s)$ on $\Omega_0^\eps\,\H$
\be{ \limp U^\eps(t)^{-1} \ueps(t + s) \Omega_0^\eps = \veps(s)\,\Omega_0^\eps
\, }
\ee
with
\be{\veps(t)\, \Omze = \Omze \,U_0(t)\ \ \ \mbox{on} \ \ \H \, . }
\ee

\noindent {\bf 2}) 
{\em Convergence of the M{\o}ller operators when $\eps \to 0$}. Namely, 
$\Omze(g)$ converges strongly to an isometric operator 
$\Omz (g) $, as $\eps \to 0$; 
then, by  eq.\,(4.3),
\be{ \lim_{\eps \to 0} \veps(s) \,\Omega_0^\eps  =
\lim_{\eps \to 0} \veps(s) \,\Omega_0  =  \vs \,\Omega_0 \, ,}
\ee
\be{ \vs \,\Omega_0 = \Omega_0\, U_0(s)\, ,}\ee
with $\vs$ a one-parameter unitary group on $\Omega_0^\eps \,\H$
which coincides with $U(s)$ under the following assumption.

\noindent {\bf 3}) 
{\em Stability of the limit $\eps \to 0$ under a change of 
the switching of order 
$\eps$}. 
Actually, only independence on the choice of the origin of time is needed, 
$g e^{- \eps |t|} \to g e^{- \eps |t+t_0|}$, $t_0 \in \dR$, i.e.,
\be {\Omze(g e^{-\eps t_0}) \longrightarrow_{\eps \ra 0} 
 \Omz(g) }  
\ee

\noindent
strongly. Eq.\,(4.7) implies
$$ U^\eps(g e^{-\eps s},t)^{-1} \,  U^\eps (g,t) \, \Omze(g)  $$
$$ = U^\eps(g e^{-\eps s},t)^{-1} \, U_0(t)\, ( U^\eps(g,t)^{-1} U_0(t))^* 
\, \Omze(g) $$
\be{ \rightarrow_{t \to \infty} \Omze(g e^{-\eps s}) \, \Omz^{\eps \,*} \, \Omze
\rightarrow_{\eps \to 0} \Omz(g) \, .}
\ee
Then, since by eq.\,(4.3) and ii) 
$$ \veps(s) \, \Omze = \lim_{t \to \infty} U^\eps(t-s)^{-1} \, U^\eps(t)) \, \Omze  $$
$$ = \lim_{t \to \infty} U^\eps(g,-s)^{-1} \, U^\eps( g e^{\eps s},t)^{-1}  \, 
U^\eps(t)) \, \Omze(g) \, ,$$
one has, by eqs.\,(4.5),(4.8),
\be { \vs \, \Omz  = \lim_{\eps \to 0} \veps (s) \, \Omze = U(s) \Omz \, .}
\ee
Therefore, eq.\,(4.5) gives the standard interpolation formula
\be { U(s) \, \Omz = \Omz \, U_0(s)\, .}
\ee

The validity of eq.\,(4.7) and, more generally, of eq.\,(4.10) in QFT 
models may require the introduction of mass counterterms,
as discussed in Appendix B and in Ref.\,\cite{MS}.

\def \udet {U_D^\eps(t)}
\def \we {W^{\eps}}

\vspace{2mm} In presence  of long range  interactions, 
$\Omze$ does not converge 
and one has to combine  the adiabatic switching with the use of a suitable 
reference dynamics. 

In this case, also the definition of $U_D(t)$ 
requires an adiabatic switching, which may be performed by replacing
$U_D(t)$ is by unitary operators $\uepsdt$, satisfying 
$$i d U^\eps_D(t)/d t = (H_0 + H_{I D}(t, e^{ - \eps |t|} g))
\, U^\eps_D(t),\,\, \,\,\,U^\eps_D(0) = \id \, ,$$
 such that the following strong limits exist
\be{ s-\limpm \uepst^{-1} \uepsdt \eqq \Omepm, }
\ee 
\be{ s-\lim_{\eps \ra 0} \Omepm \eqq \Omega_\pm.}
\ee \goodbreak
In particular, for Coulomb scattering, as we shall see 
in Section 5.2, eqs.\,(4.2), (4.3) hold if $U_0(t)$ is replaced by $\uepsdt$ 
defined 
(for $|t|$ large) by   
\be{i \frac{d \uepsdt}{ d t} = (H_0 + \frac{\a  m}
{ |\p|\, |t|} e^{-\eps |t|}) \uepsdt.}
\ee

In general, we shall consider an adiabatic switching $\uepst$, 
satisfying i), ii) and eq.\,(4.2) and a family of unitary 
operators $\udet$ for which we assume:

\noindent {\bf 1}) 
For all $\eps > 0$, the following limits exist and define unitary operators: 
\be{s-\limp \udet^{-1}\,U_0(t) = V^\eps, \,\,\,\,
V^{\eps}  \,\,\,\mbox{unitary operators}.}
\ee

\def \uases {U_{as}^\eps(s)}

\noindent
Eqs.\,(4.2),(4.14) imply, $\forall \,\eps > 0$, 
\be{\uepst^{-1} \,\udet \stackrel{s}{\rightarrow}_{t \ra \infty} \,
\Omze V^{\eps *} \eqq \Omega^\eps \, ,  }
\ee
\be{ \udet^{-1}\,U_D^\eps(t + s) \stackrel{s}{\rightarrow}_{t \ra \infty} U_{as}^\eps(s) 
= V^{\eps} U_0(s) V^{\eps *}.}\ee
Eq.\,(4.3) holds as before and eqs.\,(4.4),(4.15),(4.16) imply the following 
interpolation formula between two {\em unitary groups}
\be{\Omega^\eps \,U_{as}^\eps(s)  = \veps(s)\,\Omega^\eps.}\ee

\vspace{1mm} \noindent {\bf 2})  $\udet$ must be chosen in such a way 
that $\Omega^\eps$ converges strongly, as $\eps \ra 0$, therefore 
defining  isometric operators $\Omega$: 
\be{\Omega^\eps  \stackrel{s}{\rightarrow}_{\eps \ra 0} \Omega \, . }
\ee


\noindent {\bf 3}) 
Stability of the limit $\eps \to 0$ under a change of the switching of order 
$\eps$ as in eq.\,(4.7), namely
\be {\Omega^\eps (g, s) \eqq \lim_{t \to \infty} U^\eps(g e^{- \eps s}, t) 
\, U_D^\eps (g,t)  \stackrel{s}{\longrightarrow_{\eps \to 0}}  \Omega(g) \, , }  
\ee
the existence of the limit $t \to \infty$ following from eq.\,(4.2),(4.14) as 
for eq.\,(4.15).\goodbreak 

\noindent As before (see eqs.\,(4.8),(4.9)), eq.\,(4.19) implies
\be {\lim_{\eps \to 0} \veps(s) \, \Omega^\eps(g) = 
\lim_{\eps \to 0} U^\eps(g,-s)^{-1} \, \Omega^\eps(g, s) = 
 U(s) \, \Omega (g) \, . }
\ee
Furthermore, eq.\,(4.17) implies stability of the range of $\Omega$
under $\vs$ and therefore convergence of 
$U^\eps_{as} (s) = \Omega^{\eps \, *} \ \veps(s) \, \Omega^\eps $ and of 
$U^{\eps \, *}_{as} (s)$, as $\eps \ra 0$,  
to a strongly continuous group of unitary 
operators $U_{as}(s)$, satisfying
\be{U(s) \,\Omega = \Omega\, U_{as}(s) \, .}
\ee
Equations (4.16) also implies 
\be{U_{as}(s) = \lim_{\eps \ra 0} V^{\eps} \,U_0(s)\, V^{\eps \,*}\, .}
\ee
Equation (4.22) is close to eq.\,(2.31). Actually,
eq.\,(4.14), multiplied by its adjoint, 
$$ U_D^\eps(t + s)^{-1}\, U_0(t + s) U_0(t)^{-1} U_D^\eps (t)  
\stackrel{s}{\rightarrow}_{t \ra \infty} \id \, $$
implies that $U_D^\eps(t)$ is asymptotically free, eq.\,(2.22).

It is worthwhile to stress that, even within 
a strategy of adiabatic regularization, 
the asymptotic dynamics $ U_{as}(s)$ is a strongly 
continuous one-parameter group. 
This property is already shared by $U_{as}^\eps(s)$, 
solely as a consequence of eqs.\,(4.2), (4.14). 

The $S$-matrix is defined by $S =  \Omega_+^* \Omega_-$ and, 
if Ran $ \Omega_+ =$ Ran $\Omega_-$, $S$ is unitary and 
\be{ S = \lim_{\eps \ra 0} \Omega_+^{\eps *} \Omega_-^\eps = 
\lim_{\eps \ra 0} V_+^\eps \,S_0^\eps \,V_-^{\eps *}, \,\,\,\, 
S_0^\eps \eqq \Omega_{0 +}^{\eps *} \Omega^\eps_{0 -};}
\ee
using $T\,U^\eps(t)\,T = U^\eps(-t)$, the discussion of 
Section 2.3 applies, with $U_\pm$ defined by the limit 
$\eps \to 0$ of eq.\,(4.16).

The operators $V^\eps_\pm$ completely account  for the corrections required, 
with respect to the standard approach, as a consequence of long range 
interactions and/or infrared effects (which prevent the convergence of 
the ``cutoff'' $S$-matrix $S_0^\eps$ as $\eps \ra 0$). 
By eq.\,(4.22), they also provide the link between the asymptotic 
dynamics $U_{\pm}(t)$ and $U_0(t)$.




\section{Asymptotic fields in Coulomb scattering}
The $N$-body Coulomb scattering has been discussed in the literature 
\cite{Do1}\,\cite{Do2}\,\cite{De1}, 
but some hidden delicate points have not been 
emphasized, with the result that incorrect conclusions have appeared 
\cite{Sc2}. 

The purpose of this section is to revisit the problem with the help of the 
general discussion of the previous sections. In particular, 
we shall focus our attention on the existence of asymptotic fields and 
their space time covariance, also in view of the fact that 
Coulomb distortions play an important role in the discussion
of the asymptotic limit of fields in full quantum electrodynamics
\cite{KF}\,\cite{Zw}\,\cite{JR}.

To this purpose, we consider the field theory formulation of Coulomb 
repulsive interaction described by the (non-relativistic) Hamiltonian 
$$ H = H_0 + H_I, \,\,\,\,\,H_0 = \int d^3 p\, (\p^2/2m)\, \psi(\p)^* 
\psi(\p),$$ 
\be{ H_I = \ume e^2  \int d^3 x \,d^3 y\,\frac{1}{4 \pi |\x - \y|} 
\psi^*(\x) \psi^*(\y)\,\psi(\y) \psi(\x),}\ee
where $\psi, \psi^*$ are canonical (bosonic or fermionic) fields. 

Such a model has been discussed \cite{Sc2} in order to get information on 
the asymptotic limit of charged fields in quantum 
electrodynamics (QED) 
and the relevant point is the control of the asymptotic 
limit  of $\psi(\p,t)$.

In subsection 5.1 we shall show that

\noindent 1) asymptotic limits $\psi_{out/in}$ of the charged field 
operators are obtained by using  Dollard reference dynamics;

\noindent 2) they are space time covariant free fields ($U(t) \eqq e^{- i H t}$)
\be{ \psi_{out/in}(\p, t) \eqq  U(t)^*\,
\psi_{out/in}(\p)\,U(t) = e^{- i \p^2 t/2m} \,\psi_{out/in}(\p);}
\ee

\noindent 3) the Heisenberg $S$-matrix $S_H$ exists as a 
unitary operator, free of infrared divergences, and satisfies  
\be{ S_H\,\psi_{in} = \psi_{out}\,S_H,\,\,\,\,S_H \,U(t) = U(t)\,S_H.}
\ee

\vspace{1mm}  In subsection 5.2 the $\eps$ regularization 
of Dollard approach shall be discussed, providing 
a factorization of the infrared divergences in the $S$-matrix. 


\subsection{Heisenberg asymptotic fields}
Denoting  by $N$ the number operator, the model is defined on the Hilbert 
space $\H = \oplus_n \H^n$, $N \H^n = n \H^n$. 
In fact, the Hamiltonian $H$ commutes with  $N$, $H= \sum_n H^n$ 
with $H^n$ self adjoint on $D^n  \eqq D(H_0) \cap \H^n$, 
since $H_I$ is Kato small with respect to $H_0$ on each $D^n$; 
hence $H$ is essentially self-adjoint on $D(H) = \cup_n D^n$.

For any $n$, according to  the center of mass decomposition 
$ \H^n = L^2(\Rbf^3) \otimes L^2(\Rbf^{3(n-1)})$,  we define  
\be{ H'_n \eqq H^n - H^n_{CM}
= \id \otimes (h_{0\,n} + v_n), \,\,\,v_n(x) \eqq  e^2 
\sum_{i<j} \frac{1}{4 \pi |\x_i - \x_j|}  \, ,}
\ee 
with 
$ H^n_{CM} \eqq (\sum \p_i)^2/2 n m$, $x=\x_1, ...\x_n $;
$h_n \eqq h_{0\,n} + v_n$ is self-adjoint on 
$ D(h_{0\,n})$.

The repulsive potential implies that there is only the free particle 
channel, so that  the scattering spaces $\H^\pm$ may be identified with $\H$.

In fact, for given $n$, the channels are indexed by the point spectrum of 
$H'_{n_1}, ...H'_{n_k}$, 
for all partitions of $1, ...n$ in $k$ subsets, consisting 
of $n_i$ particles (Ref.\,\cite{De1}, Theor.6.15.1)
and, as proved below, $\forall n$, $H'_n$ has no point spectrum.
\def \hon {h_{0\,n}}
\def \dibf  {\mbox{\boldmath $\nabla$}_i}
\def \limpm {\lim_{t \ra \pm\infty}}
\newcommand{\SI}{\rm{sign}}
\begin{Proposition}  1) The Hamiltonians $H'_n$ have no point spectrum;

\noindent 2) the following strong limits exist
\be{s-\lim_{t \ra \pm\infty} U(t)^*\,\udt \eqq \Omega_\pm,}
\ee
$$\udt = e^{- i H_0 t} \exp{-i  \,{\SI} \,t \,\ln |t|\,\frac{e^2 m }
{8\pi}\int \frac{dq \, dp \, }
{ |\p -\q|} \,\psi^*(\p) \psi(\q)^* \psi(\q)\,\psi(\p)} \, ;
$$
3) $\Omega_\pm$ are unitary operators, i.e. asymptotic completeness holds;

\noindent 4)  furthermore
\be{ s-\limpm \udt^{-1} \,\udts = e^{-i H_0 s} \eqq U_0(t),}\ee 
\be{U(t) \Omega_\pm = \Omega_\pm \, U_0(t)\, .}
\ee
\end{Proposition}
\Pf\, We give a few lines proof of the absence of point spectrum of 
$H'_n$, which is somewhat hidden in the literature \cite{La1}\,\cite{La2}. 
It exploits the fact that the expectation of the infinitesimal variation
of the kinetic and potential energy under dilations always have 
the same sign, whereas their sum vanishes on eigenvectors.

\noindent Since $D^n = D(H_{CM}) \otimes D(h_{0 \, n})$, 
for any given $n$ the absence of point spectrum  of $H'_n$ is equivalent 
to the absence of eigenstates in $D^n$. 
The operator 
$$ D_R \eqq \sum_{i = 1}^n \p_i\cdot \x_i \,f_R(|x|), \,|x| = 
\sum_i |\x_i|,\,\,\,f_R(|x|) = f(|x|/R) \in \D(\Rbf), $$
$f \geq 0$,  $f(|x|) = 1$, for $|x| \leq \delta$, 
and its adjoint $D^*_R$ are well defined on $D^n$, which is 
stable  under $\x_i f_R$. 

\noindent 
We denote by $v^\eps$ the regularization of the potential $v(|x|)$
by the replacement $ |x| \to (|x|^2 + \eps^2)^{1/2} $. Then
$\sum_i \x_i \cdot \mbox{\boldmath $\nabla$}_i v_\eps \leq 0$ 
and, by the Kato estimates, $\forall \psi \in D^n$,  
$v^\eps \psi \to v \psi$, for $\eps \to 0$. Now, 
\be{ i (D_R^* \psi,\,v_\eps \psi) - i (v_\eps \psi, \,D_R \psi) = 
 \int d x \,\sum_i \mbox{\boldmath $\nabla$}_i
  v_\eps \cdot \x_i f_R \bar{\psi}(x)\,\psi(x) \leq 0.}\ee 

\noindent On the other hand, if $H'_n \psi = E \psi$, then 
$v^\eps \psi \to (E - h_{0\,n})\psi$ and the left hand side of  
eq.\,(5.8) converges to
$$ -i [(D_R^*\psi,\,h_{0 \,n}\,\psi) - (h_{0 \,n} \psi, D_R\,\psi)] \eqq \Delta H_R
\, . $$

\def \hon {h_{0\,n}}
\def \dibf  {\mbox{\boldmath $\nabla$}_i} \def \djbf  
{\mbox{\boldmath $\nabla$}_j}

\noindent 
Moreover,
$$\x_i f_R \hon \psi = \hon \x_i f_R \psi + m^{-1}[ \dibf (f_R \psi) - \sum_j   
\djbf(f_R \psi)/n] + \chi_R,$$
with $\chi_R \to 0$, for $R \to \infty$. 

\noindent Then,  since $\p_j D^n \subseteq D(\p_i)$, $\forall i,j$, 
$\forall \chi \in D^n$ one has $(\p_i \psi, \,\hon \chi) = 
(\hon \psi,\, \p_i \chi )$ and, for $R \to \infty$, 
$$ \Delta H_R \sim m^{-1} [ \sum_i (\dibf \psi, \,
\dibf \,f_R \psi ) + 
( \sum_i \dibf \psi,\, \sum_j \djbf  (f_R \psi)/n ) ] \, ,$$
which converges to $2 (\psi,\, \hon \psi) > 0 $,
contradicting eq.\,(5.8).  

\noindent The existence of $\Omega_\pm$ has been proved by Dollard 
\cite{Do2}; 
the unitarity of $\Omega_\pm$, i.e. asymptotic completeness, is a special 
case of theorem 6.15.1 of \cite{De1} (which uses the same Dollard 
modified asymptotic dynamics), since only the $N$-particle 
channel enters in eq.\,(6.15.4) of Ref.\,\cite{De1}. 

\noindent The left hand side of eq.\,(5.6) is given by 
\be{U_0(s) \exp{-i  \ln \frac{|t+s|}{|t|}\,\frac{e^2 m }
{8\pi}\int dq \, dp \, \, 
|\p -\q|^{-1} \,\psi^*(\p) \psi(\q)^* \psi(\q)\,\psi(\p)}\, .}
\ee
Such unitary operators converge strongly to $U_0(s)$  on the dense subspace 
$\cup_{n\,\eps} \H^n_\delta, $
with $\H^n_\delta \subset L^2(\Rbf^{3n})$ 
defined by $f(\p_1, ...\p_n) = 0$ whenever $|\p_i - \p_j| \leq \delta$, 
for some $i, \,j$. 
By unitarity, such strong convergence holds on $\H$. 
By Proposition 2.2, this implies the interpolation formula (5.7), 
which has also been proved by Dollard. \endproof

\vspace{2mm}
Putting $\rho(q) \eqq \psi^*(q)\psi(q)$ and denoting by
$\rho_t$, $\psi_t$ the Heisenberg fields at time $t$, we have

\begin{Proposition} 1) The Heisenberg fields 
$$ \psi_{LSZ}(\p, t) \eqq U(t)^* \udt \,\psi(\p) \,\udt^* U(t)  $$ 
\be{ =   e^{i \rho_t (C_t (\p))} \, e^{i \p^2 t/2m} 
\, \psi_t(\p)\, , }
\ee
with 
$$ \rho_t (C_t (\p)) \eqq - \, {\mathrm{sign}}\,t \,\ln |t| \,
\frac{e^2 m}{4\pi} \int dq \, \frac{\rho_t(\q)}{|\p - \q|} \,
$$ 
converge strongly on $\cup_n \H^n$, for $t \to \pm \infty$, 
after $L^2$ smearing in $\p$ 
and define asymptotic fields $\psi_{out/in}$
\be{ \psi_{out/in}(f) =  s-\limpm \psi_{LSZ}(f,t) \, , \ \ \ 
\forall f \in L^2(d^3p)\, ;}
\ee

\noindent 2) $\psi_{out/in}$ are space time covariant 
canonical free fields
\be{ \psi_{out/in}(\p, t) \eqq U(t)^*\,\psi_{out/in}(\p)\,U(t) =  
e^{- i \p^2 t/2m}\,\psi_{out/in}(\p)\, .}
\ee
and 
\be{ H = H_0(\psi_{out/in},    \psi^*_{out/in}), \,\,\,\,\mbox{on} \,\,\,D(H);}
\ee

\noindent 
3) the Heisenberg $S$-matrix $S_H$ is a unitary operator and satisfies
\be{ S_H\, \psi_{in} \,= \psi_{out}\,S_H,\,\,\,\,\,[S_H, \,U(t)\,] = 0;}
\ee
furthermore, the Schroedinger $S$-matrix 
$S = \Omega^*_+\,\Omega_-$ is unitary and commutes with $U_0(t)$
\end{Proposition}
\Pf\, The limits (5.11) exist on each $\H^n$ as a consequence of the 
unitarity of $\Omega_\pm$ (point 3 of Proposition 5.1).
Eqs.\,(5.12), (5.13), (5.14) follow by the general results of 
Section 3, with $\H_{out/in} = \H$. \endproof

\vspace{2mm}
In agreement with the discussion of Section 3.1 (eqs.\,(3.7), (3.13)),
eq.\,(5.11) provides a LSZ (HR) formula for the asymptotic limit
of the charged fields, the Dollard correction amounting to
the replacement $\psi_t(p) \to \psi_t(p) \, e^{i \rho (C_t(p)}$ 

The asymptotic fields $\psi_{out/in}$ are 
substantially different from those proposed by Zwanziger 
(\cite{Zw}, eq.\,(4))
and argued by Schweber to follow from Dollard-Kulish-Faddeev approach
(on the basis of an incorrect interpolation formula, 
eq.\,(6) in \cite{Sc2}).
A crucial property which distinguishes our $\psi_{out/in}$ from those 
proposed by  Zwanziger and later considered by Rohrlich \cite{JR} is their 
space time covariance and free time evolution, exactly as in the 
short range case.
The essential point is that Coulomb distortions do not affect
the dynamics of asymptotic fields, eq.\,(5.12) and
the Dollard modification has the role of allowing for the 
existence of the asymptotic limits through a modification of the
LSZ prescription, eq.\,(5.10).
 
\vspace{2mm}
We remark that, even if the asymptotic fields 
depend on the (arbitrary) choice of an initial time, implicit in 
the Dollard dynamics, neither their  canonical 
structure of the (anti)commutation relations nor their time evolution 
are affected. Such arbitrariness amounts to unitary tranformations,
as discussed in Section 2.1.  

 
The construction of the asymptotic variables may also be done directly 
on the observables $\x_i, \,\p_i$, according to eq.\,(3.7). 
The Dollard correction vanishes for the momenta and one has
\be{ e^{i \bbf \cdot \p_{i; out/in}} = s- \limpm U^*(t) \, 
e^{i \bbf \cdot \p_{i}}\,U(t)\, ;}\ee
in fact, the limit defines a (weakly measurable and therefore) 
strongly continuous one-parameter group. 
By  the estimates
$$ ||\frac{e^{i \bbf \cdot \p} - \id}{|\bbf|} \,U(t) \psi|| 
\leq || |\p|\,U(t) \psi|| \leq || (m H_0 + \id)\,U(t)\,\psi|| \, ,$$ 
and, by the Kato estimate $||H_0\,U(t)\,\psi || 
\leq c ||H \psi|| + b ||\psi||$, the derivative with respect to 
$(b_{i})_k$ is defined by a limit which is uniform in $t$ and one has 
\be{ \p_{i; out/in} = s-\limpm U^*(t) \p_i\,U(t), \,\,\,\,
\mbox{on}\, D(H_0). }
\ee
For the positions, since again the limit defines a strongly 
continuous one-parameter group,  one has
\be{ e^{i \abf \cdot \x_{i; out/in}} = s-\limpm U^*(t)  
e^{i \abf \cdot (\x_{i} - \p_i t/m )}\,e^{ - i\, {\mathrm{sign}} \,t \ln|t| 
[v_D(\p + \abf) - v_D(\p)]}
\,U(t)\, ,}\ee
where 
$$ v_D(\p_i) =\frac{e^2}{4 \pi} \sum_{j \neq i} \frac{1}{|\p_i - \p_j|}\, ;$$
the result is set of {\em canonical } variables 
$\x_{i; out/in}$, $\p_{i; out/in}$; their time evolution is free,
as a consequence of eq.\,(5.13), 
\be{U^*(t)\,\x_{i; out/in}\,U(t) = \x_{i; out/in}\,+\, \p_{i; out/in}\,t\, /m \, .}
\ee 


\subsection{Adiabatic procedure for Coulomb scattering}
An $\eps$ regularization combined with Dollard approach (Section 4) 
provides an explicit separation of the infrared divergences in the 
$S$-matrix and their factorization, in the case of Coulomb scattering.

The $\eps$ regularized dynamics $\uepst$ is defined  as the solution of 
equation (4.1), with $H_I$ given by eq.\,(5.1), which exists and 
is unique as a consequence of Proposition A.1 and the Kato 
smallness of $H_I$ with respect to $H_0$ on each $D^n$. Similarly, 
according to Section 4, an $\eps$ regularized  dynamics $\udet$ is defined as
\be{\udet = U_0(t) e^{- i L(\eps, t) \,V_D}, \,\,\,\,\,\,\,L(\eps,t) \eqq 
\mbox{sign}\, t \int_1^{|t|} d s \, e^{-\eps\,s}/s \, ,}
\ee
\be{V_D \eqq  \frac{e^2 m }{8\pi}\int dq \, dp \, \, |\p -\q|^{-1} \,
\psi^*(\p) \psi(\q)^* \psi(\q)\,\psi(\p)\, .}
\ee

\begin{Proposition} For $t \to \pm \infty$, 
$\uepst$ and $\udet$ satisfy eqs.\,(4.2), (4.4), (4.14), (4.15), (4.18), 
(4.20), (4.23),   
with $\Omega_\pm$ given by eq.\,(5.5).
\end{Proposition}
\Pf\, For the proof of eq.\,(4.2), we note that $\forall \psi \in D^n$
\be{|| (d/dt) \uepst^* \,U_0(t)|| = e^{-\eps |t|} ||H_I U_0(t)\,\psi|| 
\leq e^{-\eps |t|} ( c_n ||H_0 \psi|| + b_n ||\psi||)\, ,}
\ee 
which is integrable in $t$. Equation (4.4) follows.

\noindent For eq.\,(4.14), we note that $\exp[  i L(\eps, t)\,V_D]$ 
are multiplication operators which converge strongly, 
for $t \to \pm \infty$,
on $\cup_{n\, \delta} \H^n_\delta$ 
(where $|\p_i - \p_j| \geq \delta$) and therefore on $\H$, 
to the unitary multiplication operators $ V^{\eps}_\pm = 
\exp[ i L(\eps, \pm \infty)\,V_D]$. 
This implies eqs.\,(4.15), (4.16), and therefore eq.\,(4.17) 
with $U_{as}^\eps(s) = U_0(s)$, $\forall \eps > 0$.

\noindent
For eq.\,(4,19) it is enough to prove, that 
$\forall \psi \in \S_{n,\,\eps} \eqq \S(\Rbf^{3n}) \cap 
\H^n_\delta$,  $\lim_{t \to \infty} \uepst^*\,\udet \psi$ is 
uniform in $\eps$, so that the limits may be interchanged and 
one gets ($t \to \infty$ for simplicity)
\be{\lim_{\eps \to 0} \lim_{t \to + \infty} \uepst^*\,\udet \psi = 
\lim_{t \to + \infty} \lim_{\eps \to 0} \uepst^*\,\udet \psi = \Omega \psi\, .}
\ee
To prove uniformity in $\eps$ we show that
\be { ||(d/dt) \uepst^*\,\udet|| = e^{- \eps t} || (H_I - V_D/t) 
\udet \psi|| \, .  }
\ee
is majorized, independently of $\eps$, by an integrable function
of $t$. In fact, denoting 
$$ R^\eps_\psi(t) \eqq (\uepsdt - T_t e^{-i L(\eps, t)\,V_D})\,
\psi = T_t(e^{ i (m/2t) \sum_i\x_i^2} - 1)e^{-i L(\eps, t)\,V_D}\,\psi \, ,$$
$$T_t \eqq U_0(t) \,e^{ - i (m/2t) \sum_i\x_i^2}\, ,$$
Dollard estimates (\cite{Do2}, eqs.\,(134), (138))
$$|| R^\eps_\psi(t) || \leq c (\ln|t|)^6/ |t|, \,\,\,\,|| \,
|\x_i - \x_j|^{-1} R^\eps_\psi(t) || \leq c' (\ln|t|)^6/|t|^2$$
follow from 
$$| e^{-i L(\eps, t)\,V_D} \psi(x)| \leq c^{"} \frac{(\ln|t|)^6}
{ (1 + \sum_i \x_i^2)^3}, \,\,\,\,\,\,\frac{1}{|\x_i - \x_j|} T_t =  
T_t \frac{m}{|\p_i - \p_j|\,|t|}\, , $$
using the Kato smallness of $|\p_i - \p_j|^{-1}$ with respect to $\Delta_p$:
$$||\, |\p_i-\p_j|^{-1}\,f ||_{L^2} \leq a || 
\x^2 f ||_{L^2} + b ||f ||_{L^2} \, ;$$
then, 
\be{|| (H_I - V_D/t) \uepsdt \psi|| \leq || H_I\,R^\eps_\psi(t) || + 
|| R^\eps_{V_D \psi}(t)||/t = O((\ln|t|)^6/t^2)\, .}
\ee
Since $U^\eps_{as} = U_0$,  $\Omega^\eps \to \Omega$ 
and eq.\,(4.17) imply the strong convergence
\be{\tilde{U}^\eps(s)\,\Omega^\eps \to \Omega \,U_0(s) \, ;}
\ee
eq.\,(5.7) and unitarity of $\Omega$ imply 
$\tilde{U}^\eps(s) \to U(s)$, so that eq.\,(4.20) holds. 
\noindent 
Eq.\,(4.18) and asymptotic completeness imply eq.\,(4.23).
\endproof

\vspace{2mm}
\noindent {\bf Remark}.
Stability of the $\eps$ regularization, in the sense of condition 3) of
Section 4, also holds since a modified adiabatic switching,
$ e^2 \to e^2 \, e^{- \eps |t+s|}$, in the definition of $U^\eps (t)$ 
changes the r.h.s.
of eq.\,(5.23) only by the addition of a term $e^{-\eps t}  O(\eps)/t$,
which gives a contribution to the r.h.s. of eq.\,(5.22) bounded
by $ O(\eps) \ln 1/\eps $; then $\Omega^\eps(e^2 e^{-\eps s}) \to \Omega (e^2)$
as $\eps \to 0$. 
As in Section 4, stability of the $\eps$ regularization implies
$\tilde U^\eps (s) \to U(s)$ and therefore eq.,(5.7).

\vspace{2mm}
Eqs.\,(4.15), (4.18) allow for a representation of the $S$-matrix in the form
\be{ S = \lim_{\eps \to 0} e^{i L(\eps, \infty)\,V_D}\,S^\eps_0 
\,e^{-i L(\eps, -\infty)\,V_D}\, ,}
\ee
with 
\be{S_0^\eps \eqq \Omega^{\eps\, * }_{0\, +}\,\Omega^{\eps}_{0\, -},}\ee
the standard $S$-matrix with an adiabatic $\eps$ regularization.

Since
\be{ L(\eps, \infty) = - L(\eps, - \infty) \sim \ln \eps \, , }\ee
one has an explicit factorization of the infrared divergences, given by the 
divergent operator $e^{i \ln \eps\,V_D}$ on both sides of the standard 
$S$-matrix. Such operators are diagonal in $p$ space and reduce to 
phase factors there.

\vspace{5mm}

\newpage
\begin{center}
 {\Large \bf {Appendix}}  
\end{center}

\appendix

\section{Conditions for an asymptotically free reference dynamics}

A natural condition on a reference dynamics $\udt$ 
is that it is the solution of eq.\,(2.1) 
with  $ \delta \hdt \eqq \hdt - H_0 $ "vanishing at large times" 
(see e.g. eq.\,(2.2)).
In order to make such a condition more precise, we consider a 
concrete version of theorems by Kato \cite{Ka}:
\begin{Proposition} Let  $\hdt = H_0 + \delta \hdt$ on $D_0 \eqq D(H_0) 
\subseteq \H^\infty$, 
\be{ ||\delta \hdt \psi || \leq a(t) \, (|| H_0 \psi|| + b \, || \psi||), 
\,\,\,\,\,\,a(t) < 1,\,\,\,\,\,\forall \psi \in D_0}\ee
and  $\delta \hdt$  strongly differentiable in $t$ on $D_0$; then one has \goodbreak

\noindent  i) $\hdt$ is self-adjoint on $D_0$;

\noindent ii) there exists a unique family  of strongly continuous 
unitary operators $U_D(t,s)$, $t, s \in \Rbf$, 
satisfying $U_D(t,s) D_0 \subseteq D_0$ and 
\be{ i\, d U_D(t,s)/dt = \hdt \, U_D(t,s), \,\,\,\,\,\mbox{on}\,\,D_0;}
\ee 
\be{ U_D(t,s) U_D(s,\tau) = U_D(t,\tau).}
\ee
Moreover, one has  $$ i \,d U_D(t, s)/d s =  - U_D(t, s)\,H_D(s), \,\,\,\,on \,\,D_0.$$
\end{Proposition} 

\noindent \Pf\,i) follows from Kato theorem. ii) follows 
from \cite{Ka} for $t \geq s$. 
For $ t < s$, $U_D(t, s)$ is obtained (\cite{RS}, Theorem X.71) 
as the solution of eq.\,(A.2) 
with $\hdt$ replaced by $-\hdt$. Then, 
$$ i\, dU_D(t, s)/d s = - U_D(t, s)\,H_D(s)$$ 
and $U_D(t,s) U_D(s,\tau) = U_D(t,\tau)$ follows  $\forall t, s, \tau \in \Rbf$.
\endproof

\vspace{1mm}
Proposition A.1 provides the existence and uniqueness of the reference 
dynamics $U_D(t) \eqq U_D(t, 0)$ as the solution of eq.\,(2.1) on the 
domain $D(H_0)$ (stable under $\udt$ and $\udt^{-1}$). 
Proposition A.1 does not directly apply to the case of Coulomb scattering,
eq.\,(2.2), since $1/|p| t$ is 
not Kato small with respect to $H_0$; however, 
the replacement $1/|p|t  \ra 1/(|p|t +1)$ defines a reference dynamics 
yielding unitarily equivalent M{\o}ller operators and Proposition A.1 applies. 

\begin{Definition} 
Under the above assumptions, $\delta \hdt$ is said to be 
{\bf asymptotically vanishing} if, in eq.\,(A.1),  
$a(t) \ra 0$ for $|t| \ra \infty$.
\end{Definition}

\begin{Proposition} If $\hdt$ is asymptotically vanishing, then  
\be{ U_D(t+s, t) = \udts \,\udt^{-1} \,
\stackrel{s}{\longrightarrow}_{|t| \ra \infty} U_0(s)\, .}\ee  
Moreover, if $D_0$ is stable under $U_D(t)$ in the strong form
\be{ || H_0 \udt \psi|| \leq c \,||H_0 \psi|| + b'\, 
||\psi||, \,\,\,\,\,\forall \psi \in D_0 \, ,}
\ee
one has 
\be{ \udts - U_0(s) \,\udt  \stackrel{s}{\longrightarrow} 0}
\ee
for $|t| \ra \infty$, uniformly for $ s$ in finite intervals, i.e., 
$U_D(t)$ is asymptotically free. 
\end{Proposition}
\Pf\, In fact, $\forall \,\psi, \chi \in D_0$, $  t \to \infty$,
$$|(d/ds)(\chi, U_0(s)^* \udtst \psi)|=
|(\delta H_D(t+s) U_0(s) \chi, \udtst \psi)|\leq $$ 
$$\leq a(t+s) ( ||H_0 \chi|| + b\, ||\chi ||) \, || \psi|| \ra_{t \ra \infty} 0.$$
Therefore,  $\forall \psi \in \H^\infty$, 
$$U_0(s)^* \udtst \psi \ra_{weakly} \psi \, ;$$
actually, the convergence is strong, since the norm is preserved.

\noindent Moreover, $\forall \psi \in D_0$,
$$ || (d/ds) \udt^*  U_0(s)^* \udts \psi || = || \dhts \,\udts \psi|| \leq$$
$$\leq a(t+s) \,(|| H_0\, \udts 
\psi || + b\, ||\psi||) \ra_{|t| \ra \infty} 0 \, ,$$
by eq.\,(A.5). This implies $U(t)^{-1}_D U_0(s)^{-1} \udts \ra \id$ 
and eq.\,(A.6) follows. \endproof

Eq.\,(A.4) also reads
\be{ \udts^{-1} \sim_{|t| \ra \infty}  \udt^{-1} U_0(s)^{-1},} 
\ee
which means that $\udt^{-1}$ has $U_0(s)^{-1}$ as its (unique) 
right asymptotic group. However, this does not imply eq.\,(2.22)
and in fact the notions of right and left asymptotic group are 
rather different since, as remarked in Section 2.2, 
left asymptotic groups are not unique.

\section{Conditions for properties 1-3 of adiabatic switching}

\noindent 
The relation between  the above general definition 
of $\eps$ regularization and the standard heuristic 
formulas \cite{Sc1} may be realized by noticing that 1), 2) 
apply to potential scattering, 
with $H_0$ the free Hamiltonian. In fact:

\vspace{1mm}
\noindent 
a) Eq.\,(4.2) follows from eq.\,(4.1) if $ H_I$ is Kato small 
with respect to  $H_0$, since then
\be{ ||H_I e^{-i H_0 t} \psi || \leq a ||H_0 \psi|| + b || \psi||,}\ee
 so that
$$|| \frac{d }{dt}\uepst^{-1}\,U_0(t) \psi || \in L^1.$$
Such a condition implies that $\uepst$ is asymptotically free.

\noindent  
$\Omega_0^\eps$ is unitary if $(U^\eps(t)^{-1} \ueps(t + s))^*$ converges and 
this easily follows   if $H_I$ is bounded or if $H_I$ is Kato small with 
respect to $H_0$ and $||H_0\,\uepst \psi||$ is polynomially  bounded in $t$ 
for $\psi$ in a dense domain. 

\noindent Clearly, both eq.\,(4.2) and the unitarity of $\Omega_0^\eps$ 
hold if $e^{-\eps |t|}$ is replaced by a function $f_\eps(t)$ with compact support.

\vspace{1mm}
\noindent b) Strong convergence of $\Omze$ is necessary 
for  the $\eps$ regularization to be  useful. 
For short  range  ($O(r^{-1-\eps}$)) potentials $\Omze$ converges strongly 
to $\Omega_0$, for all $f_\eps(t)$ which converge to $1$ pointwise and 
are uniformly bounded (in this case $||H_I\,U_0(t) \psi|| \in L^1$ and 
the result follows from $\uepst \stackrel{s}{\rightarrow}_{\eps \ra 0} U(t)$  
and from the Lebesgue dominated convergence).

\noindent
In quantum field theory models, even with short range interactions, 
the $\Omega^\eps_0$ involve divergent terms of order $1/\eps$ and in 
order to get convergence one must eliminate persistent effects by 
introducing counterterms in $H_I$. 
This happens, e.g., in the Wentzel model, see \cite{Sc1} p.\,339-351,  
and in the Pauli-Fierz model \cite{Bl} in the case of
of massive photons, where one has divergent phase factors. 
In this case, the counterterm is a mass renormalization.

\vspace{1mm} \noindent 
c) For short range potentials eq.\,(4.7) holds, 
by a dominated convergence argument as in b) and the
interpolation formula (4.10) follows.

\vspace{1mm} \noindent 
d) The  considerations of  a)  apply to condition (4.14). 
In fact, if $\delta H_D^\eps(t)$ is asymptotically vanishing (Definition A.2), 
the existence of the limit in eq.\,(4.14) follows with $\we$ an 
isometric operator; 
again, the unitarity of $\we$ follows if $|| H_0 \udet \,\psi||$ 
is polynomially bounded in $t$, for $\psi$ in a dense domain.

In quantum field theory models, as remarked before, in general the convergence 
of $\Omega^\eps$ requires the introduction of counter terms in $H_I$, 
which must be taken into account in the choice of $H_{I D}$. 
This phenomenon is displayed by the model in Ref.\,\cite{MS}.

In general, one may also exploit the choice of a reference dynamics, 
with an adiabatic 
switching satisfying 1) and 2), for the construction of M{\o}ller operators 
even in presence of persistent effects due to $H_I$. 
In this case, 3) does not hold and,
if $U_{as}^\eps$ converges as $\eps \to 0$, $\veps(s) $ 
converges to a unitary 
group $\vs$ on $\Omega \H$, leading to the interpolation formula
$\vs \, \Omega =  \Omega \, U_{as} (s)$.
A trivial QFT example is provided by a mass perturbation of a free Hamiltonian,
$ H^\eps (t) = H_0 + e^{- \eps |t|} \Delta m $, choosing $H^\eps_D(t) = 
H^\eps (t) $; then $ \vs = U_{as}(s) = e^{-i H_0 t} $.
A less trivial example is provided by the model of Ref.\,\cite{MS} 
with the Coulomb potential replaced by a short range potential.


\end{document}